\documentclass{jfm}
\usepackage{graphicx}
\usepackage{natbib}
\usepackage{amsmath}
\usepackage{amssymb}
\usepackage{booktabs}
\usepackage{pstricks}
\usepackage{tabularx}
\usepackage[squaren,Gray]{SIunits}
\usepackage{epsfig}
\usepackage{psfrag}
\usepackage{color}
\usepackage[normalem]{ulem}

\shorttitle{A. Sow, R. E. Semenko, A. R. Kasimov} 
\shortauthor{A. N. Other et al} 
\title[]{On a stabilization mechanism for low velocity detonations}

\author[A. Sow, R. E. Semenko, A. R. Kasimov]%
{Aliou SOW$^1$, Roman E. Semenko$^2$
and Aslan R. Kasimov$^1$\thanks{Email address for correspondence: aslankasimov@gmail.com} }

\affiliation{$^1$Applied Mathematics and Computational Science \\
Room 4-2226, 4700 King Abdullah University of Science and Technology \\
Thuwal 23955-6900, Saudi Arabia\\
$^2$Department of Mechanics and Mathematics, Novosibirsk State University\\
630090, Pirogova St. 2, Novosibirsk, Russia\\
and Sobolev Institute of Mathematics,  630090, Acad. Koptyug Av., 4, Novosibirsk, Russia}

\begin{document}

\maketitle

\begin{abstract}
We use numerical simulations of the reactive Euler equations to analyze
the nonlinear stability of steady-state one-dimensional solutions
for gaseous detonations in the presence of both momentum and heat
losses. Our results point to a possible stabilization mechanism for
the low velocity detonations in such systems. The mechanism stems
from the existence of a one-parameter family of solutions found in \cite{semenko2016}.
\end{abstract}


\section{Introduction}

Investigation into the propagation of detonation in the presence of
momentum and heat losses began in the 1940s with the seminal work
by  \cite{Zeldovich1940} and then by  \cite{schelkin1949fast}.
Such losses arise when gaseous detonation propagates in small tubes
(e.g., \cite{zel1987detonation,manzhalei1992detonation,chan1989structures,camargo2010propagation}),
narrow channels (e.g., \cite{manzhalei1998gas,ishii2011detonation,gao2016near}),
channels with obstructions (e.g. \cite{lee1984high,lee1985turbulent,teodorczyk1989propagation,chan1989structures,teodorczyk1991structure,gao2016near}),
packed beds of inert particles (e.g., \cite{lyamin1985supersonic,lyamin1991propagation,babkin1991propagation,makris1995influence,babkin2012fast})
or tubes with porous walls \citep{radulescu2002failure}. Friction
with and heat transfer to the walls or obstacles (or the mass loss
through the porous walls) decreases the velocity of detonation from
its ideal value (i.e., there exists a velocity deficit). The mechanisms
causing this velocity deficit are not well understood. The stable
propagation of low velocity detonation (LVD) is of particular interest,
whereby propagation from near acoustic speeds to speeds of about half
the ideal velocity are observed. Detonation propagating at a velocity
above this range, but below the ideal value, is termed ``quasi-detonation''.
The regime of its propagation at a velocity near the sound speed in
the burnt gases is termed ``sonic'' or ``choking''  \citep{Lee-2008}.
Extensive efforts aimed at explaining these phenomena are reviewed in
\citep{gelfand1991gaseous,Lee-2008,ciccarelli2008flame,brailovsky2012combustion}.

The simplest theoretical analysis of detonation with losses follows
the formulation by  \cite{Zeldovich1940}, which is based
on the one-dimensional (1D) reactive Euler equations 
modified to account for momentum and heat losses. In qualitative agreement
with experiments, the analysis predicts that these losses lead to
a velocity deficit. In most previous works, a reverse $S$ shaped
dependence of the detonation velocity, $D$, on the loss coefficient,
$c_{f}$, has been derived \citep{zeldovich1960theory,zel1987detonation,brailovsky2000hydraulic-a,brailovsky2002effects,brailovsky2012combustion,higgins2012steady}.
The top branch of the $D$-$c_{f}$ curve corresponding to small velocity
deficits is generally stable and attracting. Even in the presence of hydrodynamic
instabilities, the average propagation velocity remains in the vicinity
of the top branch \citep{zhang1994friction,dionne2000transient}. The
middle branch is always found to be a repeller, such that solutions
starting near that branch tend toward the top branch or fail.
For this reason, the solutions corresponding to the middle branch
have been considered unphysical and often not computed \citep[e.g.][]{zel1987detonation,dionne2000transient}.  It should be noted that LVDs observed in experiments often propagate stably at rather low velocities \citep[e.g.][]{lyamin1991propagation} that are likely to correspond to the unstable middle branch of theoretical predictions\footnote{However, to the best of our knowledge, no such direct comparison is available in the literature.}. This apparent discrepancy raises an important question about
the physical reasons for the observed stability of the LVD. In this
connection, one could question the applicability of the 1D modeling
of this phenomenon, especially so in view of the complex multi-dimensional
flow structures observed in experiments. In this paper, we show evidence
of a mechanism that provides stability to the lower branch solutions
within 1D approximation. 

Various possible mechanisms (not necessarily mutually exclusive) have
previously been proposed to explain LVD. One involves a fast turbulent
deflagration that propagates at a speed near the sound speed in the
products. Flames at such high speeds are presumably promoted by the
large area of the flame caused by the turbulence in the flow between
the lead shock and the deflagration front \citep{lee1984high,lee1985turbulent,teodorczyk1989propagation,teodorczyk1991structure,Lee-2008,ciccarelli2008flame}.
In \cite{manzhalei1992detonation,manzhalei1999low}, it is suggested
that the two-dimensional structure of the flame plays an important
role in providing the necessary level of coupling between the lead
shock and the flame. Another mechanism assigns a key role to frictional
heating of the gas between the lead shock and the deflagration wave
(see e.g., \cite{babkin1991propagation,korzhavin1999one,babkin2012fast}
and the review \cite{brailovsky2012combustion}) while the multi-dimensional
and turbulent phenomena are assumed to contribute in some averaged
sense and be represented through the momentum and heat loss coefficients.

As mentioned, the 1D predictions appear to contradict the observed
stability of LVD because of the unconditional instability of the middle
branch of the $D$-$c_{f}$ curve. A recent analysis of gaseous detonation
in a porous medium with both momentum and heat losses shows that instead
of the reversed $S$ shaped dependence, the set of solutions in the
$D$-$c_{f}$ plane contains a region filled with a continuum of solutions
at low velocities \citep{semenko2016} (see Fig. \ref{fig:D-cf}).
For a given mixture and fixed $c_{f}$, there is a finite and continuous
range of detonation velocities at which steady solutions exist. This
peculiarity of the solutions points to an intriguing possibility of
the existence of a stable detonation in a wide range of low velocities.
Indeed, any particular solution inside the set-valued region is not
isolated, but is surrounded by a continuum of other steady-state
solutions. Therefore, a small perturbation to such a steady solution
can be expected to simply shift the solution to a neighboring one.
To illustrate this picture, one could draw the following mechanical
analogy with a particle moving in a potential energy field: the top
and the middle branches of the reversed $S$ shaped dependence are
analogous to a potential minimum (stable) and maximum (unstable),
respectively. The set-valued region is then analogous to a \emph{flat}
maximum, and hence it is a neutrally stable equilibrium. Any solution
that starts in the set-valued region can be expected to remain in
the neighborhood of its initial condition for a long time. That this is indeed the case is demonstrated
by the computations below.

The existence of (at least neutrally) stable solutions at low velocities
is investigated by solving the reactive Euler equations including
momentum and heat losses with initial conditions taken as the steady
traveling wave solutions of the system. In other words, we analyze
the nonlinear stability of solutions obtained in \cite{semenko2016}.
Our main finding is that the steady solutions inside the set-valued
region are long-lived as expected and either transition to the top-branch
pulsating regimes or fail over very long times, confirming our expectation
of increased stability of low velocity detonations. We show that during
the transition from LVD to pulsating solutions, the lead shock and
the deflagration front propagate at nearly the same speed in an almost steady
manner for a substantial length of time before a rapid transition
takes place.

The remainder of the paper is organized as follows. In Section \ref{sec:Governing-equations},
we introduce the reactive Euler equations that include momentum and
heat exchange terms and briefly explain the numerical algorithm. In
Section \ref{sec:Steady-state-solution}, we present the steady state
solutions and in Section \ref{sec:Unsteady-numerics}, we present
the main computational results. In Section \ref{sec:Mechanism}, we discuss the mechanism of the initial slow acceleration of the wave and its transition to the upper-branch solutions. Concluding remarks are offered in
Section \ref{sec:Conclusions}.

\section{\label{sec:Governing-equations}Governing equations and the numerical
method}

We consider a model describing a reactive perfect gas filling the
interstitial space of a packed bed of solid inert particles. The heat
release is assumed to follow a one-step global mechanism and the losses
to be described by algebraic functions $f$ for momentum and $h$
for heat. Then, the reactive Euler equations take the form \citep{semenko2016}
\begin{alignat}{2}
\rho_{t}+(\rho u)_{x} & = & 0,\nonumber \\
(\rho u)_{t}+(\rho u^{2}+p)_{x} & = & -f/\phi,\nonumber \\
(\rho\mathcal{E})_{t}+(\rho u\mathcal{E}+pu)_{x} & = & -h/\phi,\label{eq:REE}\\
(\rho\lambda)_{t}+(\rho u\lambda)_{x} & = & \rho\omega,\nonumber 
\end{alignat}
where $\rho$, $u$, $p$ and $\lambda$ are the density, flow velocity,
pressure and reaction progress variable, respectively, while $\phi$ denotes the medium porosity. Variable $\lambda$
varies between $0$ in reactants to $1$ in products. The total energy
density is $\mathcal{E}=p/\left(\rho\left(\gamma-1\right)\right)+u^{2}/2-\lambda Q$
with $Q$ denoting the heat release and $\gamma$ the ratio of specific
heats. The ideal gas equation of state is assumed. The reaction rate
is of the Arrhenius form, ${\displaystyle \omega=k\left(1-\lambda\right)\exp\left(-E/RT\right)}$,
with $k$ denoting the rate constant, $R$ the universal gas constant
and $E$ the activation energy. The governing equations are rescaled
with respect to the upstream state, $\rho_{a},$ $p_{a}$ and $u_{a}=\sqrt{p_{a}/\rho_{a}}$.
The length scale is chosen to be the half-reaction zone length, $l_{1/2}$,
of the ideal planar detonation, and the time scale is $l_{1/2}/u_{a}$.
The drag force and heat loss terms are assumed to follow the same modeling
assumptions as in \cite{semenko2016} and are taken as $f=c_{f}\rho u\left|u\right|$
and $h=c_{h}\left|u\right|\left(T-1\right)$. Here, the dimensionless
coefficients $c_{f}$ and $c_{h}$ depend on a large number of physical
parameters that describe the gas transport properties, the porous
medium and the chemical reactions, as the reader can find in \cite{semenko2016}. In particular, using  the assumptions of \cite{semenko2016}, that the Reynolds number is large and taking $\phi=0.4$, we obtain that 
$c_{f} = 2.7\, l_{1/2}/d$, where $d$ is the particle diameter.
Following the same reference, it is assumed that $c_{h}=0.4c_{f}$
and therefore $c_{f}$ remains the only parameter characterizing the
losses. 

For the numerical solution of the above system, we have implemented
a shock-fitting algorithm based on a fifth-order finite difference
WENO5M (Weighted Essentially Non-Oscillatory) scheme for the spatial
discretization and a fifth-order Runge-Kutta method for the temporal
integration \citep{HenrickAslamPowers2006}. As in \cite{HenrickAslamPowers2006},
the shock change equation was derived and discretized to compute the
detonation velocity. However, unlike \cite{HenrickAslamPowers2006},
biased approximations of spatial derivatives near the shock were not
used to avoid numerical stability problems when secondary shocks that
frequently arise in the reaction zone interact with the lead shock.
The lead shock state was extrapolated ahead of the detonation shock
whenever necessary, which had a drawback of reducing the order of
the method to first, but gave the advantage of a numerical stability
under a wide range of conditions. The spatial discretization of the
physical domain was based on a uniform Cartesian grid.

A particular challenge for the present computations is related to
the extreme range of spatial scales often present in simulations and
very long computational times that are often necessary. Parallel computer
resources were required in most cases, especially in the low velocity
regimes. Up to $200$ processor cores were used. Numerical resolutions
were $N_{1/2}=50$, $N_{1/2}=100$ or $N_{1/2}=200$, where $N_{1/2}$
denotes the number of grid points per unit length. The domain lengths
used were between $L=100$ and $L=4000$ depending on a particular
case. At the downstream boundary, the flow variables were extrapolated.
We assured that neither the resolution nor the domain length had an
appreciable effect on the results.

\section{\label{sec:Steady-state-solution}Steady-state solutions}

\begin{figure}{}
\centering
\psfragscanon
\psfrag{Cf}[l][Bl][0.9][0]{$c_f$}
\psfrag{D/DCJ}[][][0.9][0]{$D/D_{\textrm{CJ}}$}
\psfrag{(a)}[][][0.8][0]{(a)}
\psfrag{(b)}[][][0.8][0]{(b)}
\psfrag{a}[][][0.8][0]{a}
\psfrag{b}[][][0.8][0]{b}
\psfrag{c}[][][0.8][0]{c}
\psfrag{d}[][][0.8][0]{d}
\psfrag{e}[][][0.8][0]{e}
\psfrag{f}[][][0.8][0]{f}
\psfrag{g}[][][0.8][0]{g}
\psfrag{h}[][][0.8][0]{h}
\psfrag{i}[][][0.8][0]{i}
\psfrag{j}[][][0.8][0]{\hspace{0.2cm}j}
\psfrag{k}[][][0.8][0]{k}
\psfrag{l}[][][0.8][0]{l}
\psfrag{m}[][][0.8][0]{m}
\psfrag{0}[][][0.7][0]{$0$}
\psfrag{0.005}[][][0.7][0]{$0.005$}
\psfrag{0.01}[][][0.7][0]{$0.01$}
\psfrag{0.015}[][][0.7][0]{$0.015$}
\psfrag{0.02}[][][0.7][0]{$0.02$}
\psfrag{0.025}[][][0.7][0]{$0.025$}
\psfrag{0.03}[][][0.7][0]{$0.03$}
\psfrag{0.04}[][][0.7][0]{$0.04$}
\psfrag{0.2}[][][0.7][0]{$0.2$}
\psfrag{0.4}[][][0.7][0]{$0.4$}
\psfrag{0.6}[][][0.7][0]{$0.6$}
\psfrag{0.8}[][][0.7][0]{$0.8$}
\psfrag{1}[][][0.7][0]{$1$}
\includegraphics[scale=0.45,angle=0]{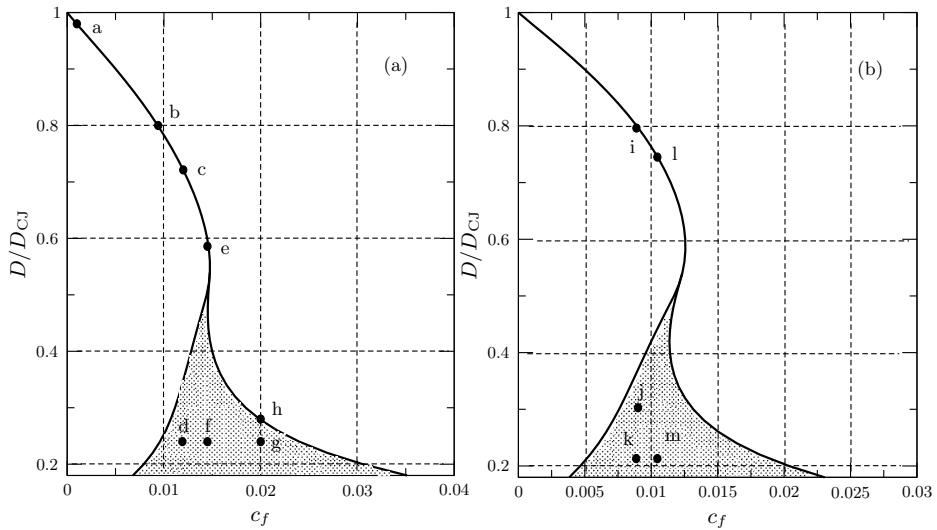}
\caption{Dependence of the detonation speed on the loss factor at (a)  $E=25$ and (b) $E=27 $. In both cases, $Q=50$, $\gamma =1.2$ and $c_h=0.4c_f$.  
 }
\label{fig:D-cf}.
\end{figure}

Steady traveling wave solutions of \eqref{eq:REE} were thoroughly
investigated in \cite{semenko2016}. It was found that solutions exist
with wave speeds, $D$, that can vary from the ideal value, $D_{\mathrm{{CJ}}}$,
down to the ambient sound speed, $c_{a}=\sqrt{\gamma}$. At each wave
speed, the solution exists only at a particular value (or range) of
$c_{f}$. These results are consistent with previous findings (e.g.,
\cite{zeldovich1960theory,zel1987detonation,brailovsky2002effects}).
However, a peculiar feature of the $D$-$c_{f}$ dependence, revealed
for the first time in \cite{semenko2016}, is the presence of a set-valued
region, i.e., a continuous range of $D$ for a given $c_{f}$ at sufficiently
low wave speeds, shown in Fig. \ref{fig:D-cf} as a shaded region.
These solutions exist only with both momentum and heat losses and
correspond to the post-shock flow without a sonic locus. They are
also accompanied by a flow reversal effect such that in some part
of the postshock flow, the gas moves opposite to the direction of
the wave in the laboratory frame of reference. Two examples of $D$-$c_{f}$
dependence at two different activation energies are shown in Fig.
\ref{fig:D-cf}. The steady-state solutions exist at every pair of
($D$, $c_{f}$) on the curves as well as inside the shaded regions.
In Table \ref{tab:D-cf}, we list the coordinates of the labeled points
in Fig. \ref{fig:D-cf}. Stability of solutions corresponding to these
points is investigated below.

 \begin{table}
  \begin{center}
  \begin{tabular}{llllllllllllllllllllllllllllllll}

     Operating points & $a$& $b$& $c$& $d$& $e$& $f$& $g$& $h$& $i$& $j$& $k$& $l$& $m$  \\[3pt]

       $D/D_{\textrm{CJ}}$   &$0.98$ &$0.8$ &$0.72$ &$0.24$ &$0.59$ &$0.24$ &$0.24$ &$0.28$&$0.8$ &$0.3$ &$0.21$ &$0.74$ &$0.21$ \\
       $c_f\times10^{-3} $   & $0.97$ & $9.37$ &$12$ &$12$ &$14.6$ &$14.6$ &$20$ &$20$ & $8.9$ & $8.9$ &$8.9$ &$10.05$ &$10.05$\\     

  \end{tabular}
     \caption{Values of $D/D_{\textrm{CJ}}$  and $c_f$ at points indicated in Fig. \ref{fig:D-cf}. } 
\label{tab:D-cf}
  \end{center}
\end{table}

\begin{figure}{}
\centering
\psfragscanon
\psfrag{t}[l][Bl][1][0]{$t$}
\psfrag{u}[][][0.8][0]{$u$}
\psfrag{omega}[][][1][0]{$\omega$}
\psfrag{x}[][][1][0]{$x$}
\psfrag{(a)}[][][1][0]{(a)}
\psfrag{(b)}[][][1][0]{(b)}
\psfrag{point c}[][][0.8][0]{\hspace{-0.4cm}point c}
\psfrag{point d}[][][0.8][0]{point d}
\psfrag{0}[][][0.7][0]{$0$}
\psfrag{1.4}[][][0.7][0]{$1.4$}
\psfrag{1.2}[][][0.7][0]{$1.2$}
\psfrag{1.04}[][][0.7][0]{$1.04$}
\psfrag{1}[][][0.7][0]{$1$}
\psfrag{0.96}[][][0.7][0]{$0.96$}
\psfrag{0.92}[][][0.7][0]{$0.92$}
\psfrag{-100}[][][0.7][0]{$-100$}
\psfrag{-200}[][][0.7][0]{$-200$}
\psfrag{-1000}[][][0.7][0]{$-1000$}
\psfrag{-1500}[][][0.7][0]{$-1500$}
\psfrag{-2000}[][][0.7][0]{$-2000$}
\psfrag{-3000}[][][0.7][0]{$-3000$}
\psfrag{0.2}[][][0.7][0]{$0.2$}
\psfrag{0.4}[][][0.7][0]{$0.4$}
\psfrag{0.6}[][][0.7][0]{$0.6$}
\psfrag{0.8}[][][0.7][0]{$0.8$}
\psfrag{1.8}[][][0.7][0]{$1.8$}
\psfrag{2.4}[][][0.7][0]{$2.4$}
\psfrag{3}[][][0.7][0]{$3$}
\psfrag{3.6}[][][0.7][0]{$3.6$}
\psfrag{4.2}[][][0.7][0]{$4.2$}
\psfrag{4.8}[][][0.7][0]{$4.8$}
\includegraphics[scale=0.45,angle=0]{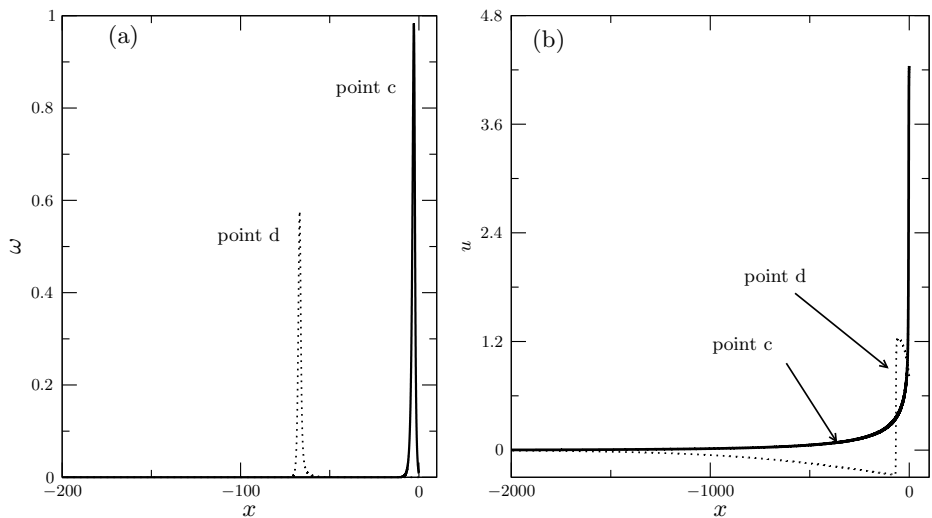}
\caption{Profiles of the steady-state reaction rate
(a) and fluid velocity (b) at points $c$ and $d$ on Fig. \ref{fig:D-cf}(a).  
 }
\label{fig:steady-states}
\end{figure}

To contrast the nature of the steady-state solutions that coexist
at the same value of $c_{f}$, we show in Fig. \ref{fig:steady-states}
the profiles of the reaction rate and flow velocity at points $c$
and $d$. Note that the velocity profile at $d$, the point inside
the set-valued region, has a substantial negative phase. As explained
in \cite{semenko2016}, such flow reversal is only possible in the
presence of heat losses. What is also important in these figures is
that the solutions in the set-valued region have a very sharp ``fire''
located far from the lead shock (Fig. \ref{fig:steady-states}(a))
and a very long velocity relaxation region downstream of the fire
(Fig. \ref{fig:steady-states}(b)). Corresponding temperature profiles
(not shown here) also exhibit striking contrast \citep{semenko2016}.
At point $c$, the shock is strong enough to raise the gas temperature
to sufficiently high levels so as to ignite the gas, and hence this structure and the wave propagation
mechanism are close to those of the ideal case. In contrast, at point
$d$, the lead shock is weak and the shock temperature is insufficient
to initiate reactions. Instead, the temperature is raised by frictional
heating in the region between the shock and the ``fire.'' A more
detailed discussion of these effects can be found in \cite{semenko2016}.

\section{\label{sec:Unsteady-numerics}Nonlinear stability results}

In this section, we explore the stability of the steady state solutions
indicated by points in Fig. \ref{fig:D-cf}. These solutions are taken
as initial conditions in the simulations. Their instability is triggered
by numerical discretization errors. Note that the ideal detonation
at $E=25$ is stable while that at $E=27$ is unstable with respect
to one-dimensional perturbations \citep{LeeStewart90}. 

\begin{figure}{}
\centering
\psfragscanon
\psfrag{t}[l][Bl][.8][0]{$t$}
\psfrag{D/DCJ}[][][0.9][0]{$D/D_{\textrm{CJ}}$}\psfrag{D}[l][Bl][0.8][0]{$D$}
\psfrag{(a)}[][][0.7][0]{(a)}
\psfrag{(b)}[][][0.7][0]{(b)}
\psfrag{point a}[][][0.7][0]{point a}
\psfrag{point b}[][][0.7][0]{point b}
\psfrag{0}[][][0.7][0]{$0$}
\psfrag{1.4}[][][0.7][0]{$1.4$}
\psfrag{1.2}[][][0.7][0]{$1.2$}
\psfrag{1.04}[][][0.7][0]{$1.04$}
\psfrag{1}[][][0.7][0]{$1$}
\psfrag{0.96}[][][0.7][0]{$0.96$}
\psfrag{0.92}[][][0.7][0]{$0.92$}
\psfrag{500}[][][0.7][0]{$500$}
\psfrag{1000}[][][0.7][0]{$1000$}
\psfrag{1500}[][][0.7][0]{$1500$}
\psfrag{2000}[][][0.7][0]{$2000$}
\psfrag{0.2}[][][0.7][0]{$0.2$}
\psfrag{0.4}[][][0.7][0]{$0.4$}
\psfrag{0.6}[][][0.7][0]{$0.6$}
\psfrag{0.8}[][][0.7][0]{$0.8$}
\includegraphics[scale=0.45,angle=0]{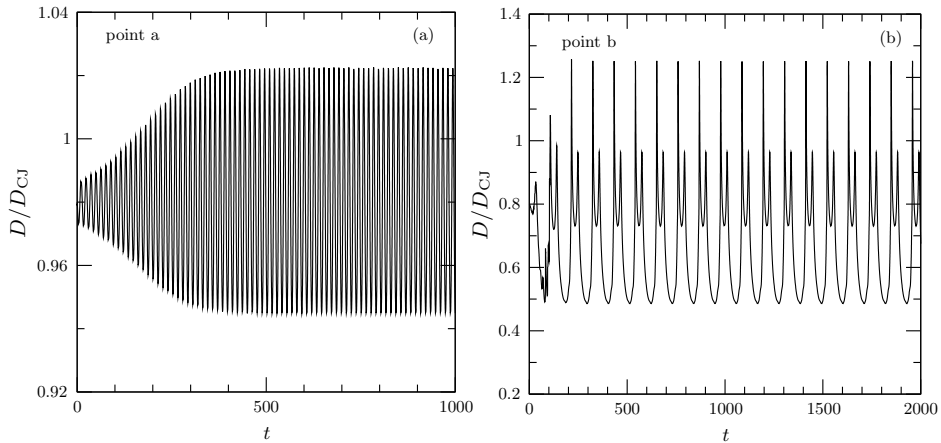}
\caption{Detonation velocity as a function
of time for solutions starting on the top branch of Fig. \ref{fig:D-cf}(a):
(a) -- point $a$; (b) -- point $b$. The period of the limit cycle
in (a) is $13.12$.
 }
\label{fig:E25-a-b-period-doubling}
\end{figure}

The presence of losses leads to the increased instability of the wave
\citep{zhang1994friction,dionne2000transient,sow2014mean}. Figure
\ref{fig:E25-a-b-period-doubling} shows the period-one and period-two
limit cycles arising from the instability of solutions at points $a$
and $b$. The first bifurcation takes place at $D/D_{\mathrm{CJ}}=0.986$,
$c_{f}=6.72\times10^{-4}$. In Fig. \ref{fig:E25-a-b-period-doubling}(a),
a simple limit cycle is seen arising because of the presence of losses.
Moving down along the top branch leads to a period-doubling bifurcation
at $D/D_{\textrm{{CJ}}}=0.909$, $c_{f}=4.57\times10^{-3}$. Figure
\ref{fig:E25-a-b-period-doubling}(b) shows the evolution of $D$
starting at point $b$. We found that the period-two solution appears
to continue down close to the turning point. 

\begin{figure}{}
\centering
\psfragscanon
\psfrag{D/DCJ}[][][0.9][0]{$D/D_{\textrm{CJ}}$}\psfrag{D}[l][Bl][0.8][0]{$D$}
\psfrag{t}[][][0.8][0]{$t$}
\psfrag{(a)}[][][0.7][0]{(a)}
\psfrag{(b)}[][][0.7][0]{(b)}
\psfrag{(c)}[][][0.7][0]{(c)}
\psfrag{(d)}[][][0.7][0]{(d)}
\psfrag{point c}[][][0.7][0]{point c}
\psfrag{point d}[][][0.7][0]{point d}
\psfrag{point e}[][][0.7][0]{point  e}
\psfrag{point f}[][][0.7][0]{\hspace{0.2cm}point f}
\psfrag{point g}[][][0.7][0]{point  g}
\psfrag{point h}[][][0.7][0]{\hspace{0.2cm}point h}
\psfrag{0}[][][0.7][0]{$0$}
\psfrag{1.4}[][][0.7][0]{$1.4$}
\psfrag{1.2}[][][0.7][0]{$1.2$}
\psfrag{1.04}[][][0.7][0]{$1.04$}
\psfrag{1}[][][0.7][0]{$1$}
\psfrag{0.96}[][][0.7][0]{$0.96$}
\psfrag{0.92}[][][0.7][0]{$0.92$}
\psfrag{500}[][][0.7][0]{$500$}
\psfrag{1000}[][][0.7][0]{$1000$}
\psfrag{1500}[][][0.7][0]{$1500$}
\psfrag{2000}[][][0.7][0]{$2000$}
\psfrag{2500}[][][0.7][0]{$2500$}
\psfrag{5000}[][][0.7][0]{$5000$}
\psfrag{7500}[][][0.7][0]{$7500$}
\psfrag{10000}[][][0.7][0]{$10000$}
\psfrag{0.1}[][][0.7][0]{$0.1$}
\psfrag{0.15}[][][0.7][0]{$0.15$}
\psfrag{0.2}[][][0.7][0]{$0.2$}
\psfrag{0.25}[][][0.7][0]{$0.25$}
\psfrag{0.3}[][][0.7][0]{$0.3$}
\psfrag{0.4}[][][0.7][0]{$0.4$}
\psfrag{0.6}[][][0.7][0]{$0.6$}
\psfrag{0.8}[][][0.7][0]{$0.8$}
\includegraphics[scale=0.45,angle=0]{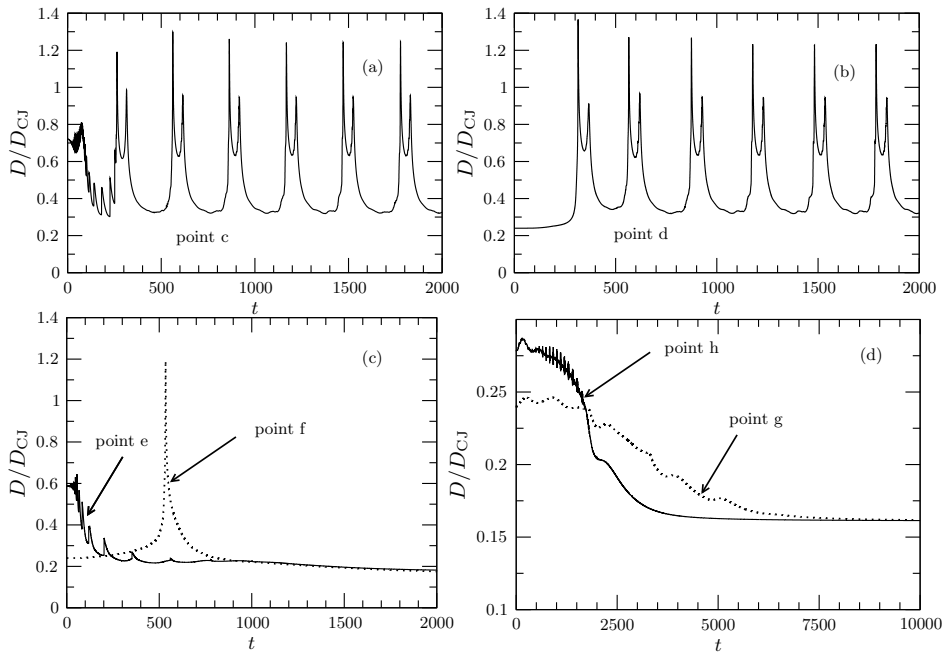}
\caption{Stability of the steady-state solutions at points $c$-$h$ of Fig. \ref{fig:D-cf}(a). 
 }
\label{fig:E25-c-h}
\end{figure}

To understand the stability properties of the set-valued solutions,
we take point $c$ on the top branch and point $d$ in the set-valued
region at the same value of $c_{f}$ and calculate the detonation
dynamics starting at these points. We find that the solution starting
at $c$ proceeds through a number of high-frequency oscillations up
until $t\approx100$ and then a rapid decay in velocity down to about
$D/D_{\mathrm{CJ}}=0.3-0.4$ is seen with a subsequent initiation
of large-amplitude low-frequency period-two oscillations around the
point $c$, as shown in Fig. \ref{fig:E25-c-h}(a). Figure \ref{fig:E25-c-h}(b)
displays a qualitatively different behavior at early times -- very
slow increase of velocity with no oscillations. At about $t=300$,
we observe that the solution transitions to the same limit cycle as
in Fig. \ref{fig:E25-c-h}(a), indicating that the trajectory is attracted
to the top branch. The long-time behavior of solutions starting at
points $c$ and $d$ as well as other points in the set-valued region
at the same $c_{f}$ are found to be identical. A different solution
starting at the same $c_{f}$, but closer to the boundary of the set-valued
region is seen to transition to the top branch faster than the solutions
starting from further down at lower velocities. What is common to
all the solutions in the set-valued region is the presence of a nonoscillatory
nearly steady early phase. The duration of this phase depends on the
position in the set-valued region as well as on other factors, such
as the activation energy.

Next, we look at the stability of the set-valued solutions at such
$c_{f}$ that there is either no corresponding top branch or for which
the top branch is near the turning point. The results are shown in
Figs. \ref{fig:E25-c-h}(c-d). They are unstable again in both cases,
but in Fig. \ref{fig:E25-c-h}(c), we see that both points $e$ and
$f$ lead to detonation failure, even though the precise path is quite
different in these cases. The set-valued point $f$ leads to a long
nearly steady solution that evolves to an explosion at about $t=500$
with a subsequent failure of the wave. Note that at this point, the
transition to the upper branch toward point $e$ takes place over
a large set-valued region above $f$. Although the wave speeds up
at the beginning as if attempting to make the transition, because
of the proximity of the upper-branch solution to the turning point,
the solution is strongly unstable resulting in the detonation failure.

Now we look at the points $g$ and $h$ for which there is no corresponding
top branch (Fig. \ref{fig:E25-c-h}(d)). In both cases, detonation
fails. However, note that the decay is very slow initially, the wave
remaining near its initial velocity for about $t=1500-2000$. The
sharp oscillations seen for point $h$ early on are the consequence
of local explosions overtaking the lead shock, which are similar to
those observed at the transition in Fig. \ref{fig:E25-c-h}(a). Their
extremely sharp appearance in Fig. \ref{fig:E25-c-h}(d) is due to
the long scales of the time axis.

\begin{figure}{}
\centering
\psfragscanon
\psfrag{D/DCJ}[][][0.9][0]{$D/D_{\textrm{CJ}}$}\psfrag{D}[l][Bl][0.8][0]{$D$}
\psfrag{t}[][][0.8][0]{$t$}
\psfrag{(a)}[][][0.7][0]{(a)}
\psfrag{(b)}[][][0.7][0]{(b)}
\psfrag{(c)}[][][0.7][0]{(c)}
\psfrag{(d)}[][][0.7][0]{(d)}
\psfrag{point l}[][][0.7][0]{l}
\psfrag{point m}[][][0.7][0]{m}
\psfrag{i}[][][0.7][0]{i}
\psfrag{j}[][][0.7][0]{j}
\psfrag{k}[][][0.7][0]{k}
\psfrag{0}[][][0.7][0]{$0$}
\psfrag{1.6}[][][0.7][0]{$1.6$}
\psfrag{1.4}[][][0.7][0]{$1.4$}
\psfrag{1.2}[][][0.7][0]{$1.2$}
\psfrag{1.04}[][][0.7][0]{$1.04$}
\psfrag{1}[][][0.7][0]{$1$}
\psfrag{2}[][][0.7][0]{$2$}
\psfrag{1.5}[][][0.7][0]{$1.5$}
\psfrag{0.5}[][][0.7][0]{$0.5$}
\psfrag{0.96}[][][0.7][0]{$0.96$}
\psfrag{0.92}[][][0.7][0]{$0.92$}
\psfrag{500}[][][0.7][0]{$500$}
\psfrag{1000}[][][0.7][0]{$1000$}
\psfrag{1500}[][][0.7][0]{$1500$}
\psfrag{2000}[][][0.7][0]{$2000$}
\psfrag{3000}[][][0.7][0]{$3000$}
\psfrag{4000}[][][0.7][0]{$4000$}
\psfrag{5000}[][][0.7][0]{$5000$}
\psfrag{6000}[][][0.7][0]{$6000$}
\psfrag{7000}[][][0.7][0]{$7000$}
\psfrag{8000}[][][0.7][0]{$8000$}
\psfrag{10000}[][][0.7][0]{$10000$}
\psfrag{0.1}[][][0.7][0]{$0.1$}
\psfrag{0.15}[][][0.7][0]{$0.15$}
\psfrag{0.2}[][][0.7][0]{$0.2$}
\psfrag{0.25}[][][0.7][0]{$0.25$}
\psfrag{0.3}[][][0.7][0]{$0.3$}
\psfrag{0.4}[][][0.7][0]{$0.4$}
\psfrag{0.6}[][][0.7][0]{$0.6$}
\psfrag{0.8}[][][0.7][0]{$0.8$}

\includegraphics[scale=0.45,angle=0]{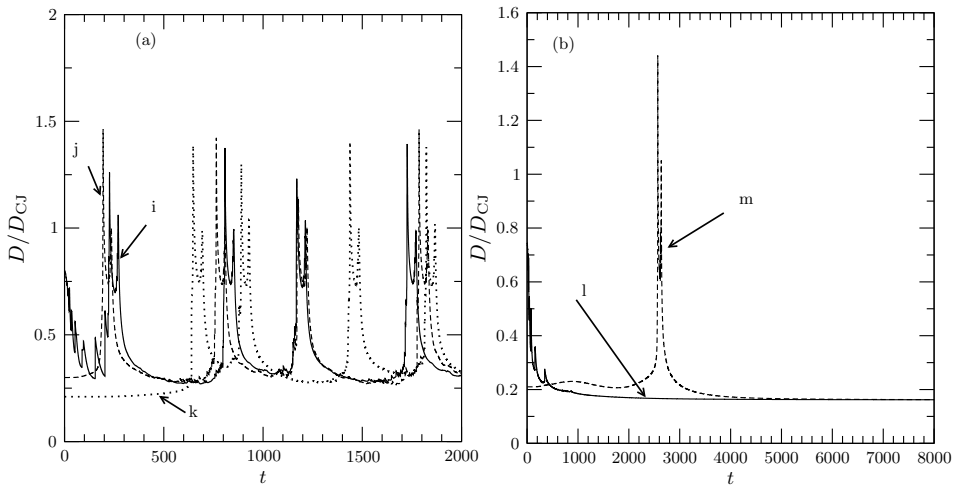}
\caption{Stability of the steady-state solutions at points $i$-$m$ of Fig. \ref{fig:D-cf}(b).
 }
\label{fig:E27-i-m}
\end{figure}

Figure \ref{fig:E27-i-m} displays similar results for a larger activation energy, $E=27$; all other parameters remain the same as before. In
Fig. \ref{fig:E27-i-m}(a), case $k$, the solution appears to be essentially steady until $t\approx600$, at which point we see a rapid
transition to the top branch. The dynamics of the solutions starting at $j$ are similar to those starting at $k$, but with a shorter nearly steady stage. Comparison with the lower activation energy case indicates that increasing the activation energy leads to a more extended
in time  initial dynamics. Figure \ref{fig:E27-i-m}(b) demonstrates a behavior similar to that at the lower activation energy of Fig. \ref{fig:E25-c-h}(c), but with a much more extended early transient stage, now continuing to around $t=2500$. 

\begin{figure}{}
\centering
\psfragscanon
\psfrag{D/DCJ}[][][0.9][0]{$\;D/D_{\textrm{CJ}}\;\;$}
\psfrag{t}[][][1][0]{$t$}
\psfrag{flame speed}[][][1][0]{fire speed}
\psfrag{wave speed}[][][1][0]{shock speed}
\psfrag{0}[][][0.7][0]{$0$}
\psfrag{1.6}[][][0.7][0]{$1.6$}
\psfrag{1.4}[][][0.7][0]{$1.4$}
\psfrag{1.2}[][][0.7][0]{$1.2$}
\psfrag{1.04}[][][0.7][0]{$1.04$}
\psfrag{1}[][][0.7][0]{$1$}
\psfrag{0.96}[][][0.7][0]{$0.96$}
\psfrag{0.92}[][][0.7][0]{$0.92$}
\psfrag{100}[][][0.7][0]{$100$}
\psfrag{200}[][][0.7][0]{$200$}
\psfrag{300}[][][0.7][0]{$300$}
\psfrag{400}[][][0.7][0]{$400$}
\psfrag{500}[][][0.7][0]{$500$}
\psfrag{600}[][][0.7][0]{$600$}
\psfrag{700}[][][0.7][0]{$700$}
\psfrag{800}[][][0.7][0]{$800$}
\psfrag{10000}[][][0.7][0]{$10000$}
\psfrag{0.1}[][][0.7][0]{$0.1$}
\psfrag{0.15}[][][0.7][0]{$0.15$}
\psfrag{0.2}[][][0.7][0]{$0.2$}
\psfrag{0.25}[][][0.7][0]{$0.25$}
\psfrag{0.3}[][][0.7][0]{$0.3$}
\psfrag{0.4}[][][0.7][0]{$0.4$}
\psfrag{0.6}[][][0.7][0]{$0.6$}
\psfrag{0.8}[][][0.7][0]{$0.8$}

\includegraphics[scale=0.35,angle=0]{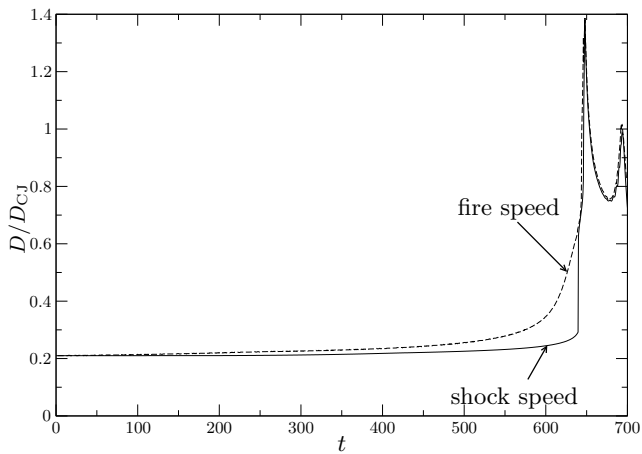}
\caption{Low velocity detonation at point $k$ of Fig. \ref{fig:D-cf}(b) and its transition to the top branch. The speeds of the detonation and the deflagration waves as a function of time are shown.
 }
\label{fig:E27-k-LVD}
\end{figure}

In Fig. \ref{fig:E27-k-LVD}, we show the velocities of the lead shock and the deflagration wave for the case of point $k$ shown in Fig. \ref{fig:E27-i-m}(a). The two velocities appear to essentially coincide. The deflagration speed slightly exceeds the shock speed up until around $t=500-600$. Then, the deflagration wave is seen to rapidly accelerate, forcing the lead shock to quickly transition to the top branch. After the transition, both fronts become closely coupled, with a much smaller distance between them than before the transition, and with similar velocities.

\section{\label{sec:Mechanism}The transition process}

Here, we describe in some detail the process of the transition from the low velocity to the high velocity branch. The cases  of successful transition (at $D/D_{\textrm{CJ}}=0.3$, which is in the set-valued region just above point $d$ in Fig. \ref{fig:D-cf}) and failure (point $g$) are considered. To elucidate the role played by different physical processes in the mechanism of the transition, we analyse various terms in the pressure evolution equation (written in the shock-attached frame) 

\begin{alignat}{2}
p_{t} & = & \underbrace{(D-u) p_{x}}_{1}\,  \underbrace{- \gamma p u_{x}}_{2} + \underbrace{(\gamma-1)Q\rho\omega}_{3} + \underbrace{(\gamma-1)\frac{uf}{\phi}}_{4} \,\, \underbrace{-(\gamma-1)\frac{h}{\phi}}_{5}.\label{eq:pressure}
\end{alignat}

This equation can be easily derived from the governing system \eqref{eq:REE}. In the steady state, all the terms on the right-hand side of \eqref{eq:pressure} balance to yield $p_t=0$. Breaking of this balance leads to the local pressure change. The mechanism of the shock acceleration from the slow early regime to the fast pulsating long term regimes is controlled by how different terms in \eqref{eq:pressure} evolve in time and space. These terms have transparent physical meaning: (1) $(D-u )p_x$ is the rate of pressure rise due to the advection of nearby gas to the given location; (2) $- \gamma p u_{x}$ is due to the compressibility of the gas; (3) $(\gamma-1)Q\rho\omega$ is due to the chemical heat release; (4) $(\gamma-1)uf/{\phi}$ is the contribution by friction; and (5) $-(\gamma-1)h/{\phi}$ is due to the heat loss.

\begin{figure}
\centering
\psfragscanon
\psfrag{(a)}[l][Bl][0.7][0]{(a)}
\psfrag{(b)}[l][Bl][0.7][0]{(b)}
\psfrag{(c)}[l][Bl][0.7][0]{(c)}
\psfrag{(d)}[l][Bl][0.7][0]{(d)}
\psfrag{x}[l][Bl][0.8][0]{$x$}
\psfrag{t=0.5}[l][Bl][0.8][0]{$t=0.5$}
\psfrag{t=140}[l][Bl][0.8][0]{$t=140$}
\psfrag{t=160}[l][Bl][0.8][0]{$t=160$}
\psfrag{t=167}[l][Bl][0.8][0]{$t=167$}
\psfrag{x}[l][Bl][0.8][0]{$x$}
\psfrag{0}[][][0.9][0]{$0$}
\psfrag{10}[][][0.9][0]{$10$}
\psfrag{12}[][][0.9][0]{$12$}
\psfrag{20}[][][0.9][0]{$20$}
\psfrag{30}[][][0.9][0]{$30$}
\psfrag{40}[][][0.9][0]{$40$}
\psfrag{50}[][][0.9][0]{$50$}
\psfrag{60}[][][0.9][0]{$60$}
\psfrag{-4}[][][0.9][0]{$-4$}
\psfrag{-8}[][][0.9][0]{$-8$}
\psfrag{-12}[][][0.9][0]{$-12$}
\psfrag{-16}[][][0.9][0]{$-16$}
\psfrag{4}[][][0.9][0]{$4$}
\psfrag{8}[][][0.9][0]{$8$}
\psfrag{1}[][][0.9][0]{$1$}
\psfrag{-50}[][][0.9][0]{$-50$}
\psfrag{-10}[][][0.9][0]{$-10$}
\psfrag{-20}[][][0.9][0]{$-20$}
\psfrag{-30}[][][0.9][0]{$-30$}
\psfrag{-40}[][][0.9][0]{$-40$}
\psfrag{-50}[][][0.9][0]{$-50$}
\psfrag{-60}[][][0.9][0]{$-60$}
\psfrag{-80}[][][0.9][0]{$-80$}
\psfrag{-100}[][][0.9][0]{$-100$}
\psfrag{-680}[][][0.9][0]{$-680$}
\psfrag{-690}[][][0.9][0]{$-690$}
\psfrag{-700}[][][0.9][0]{$-700$}\includegraphics[scale=0.5,angle=0]{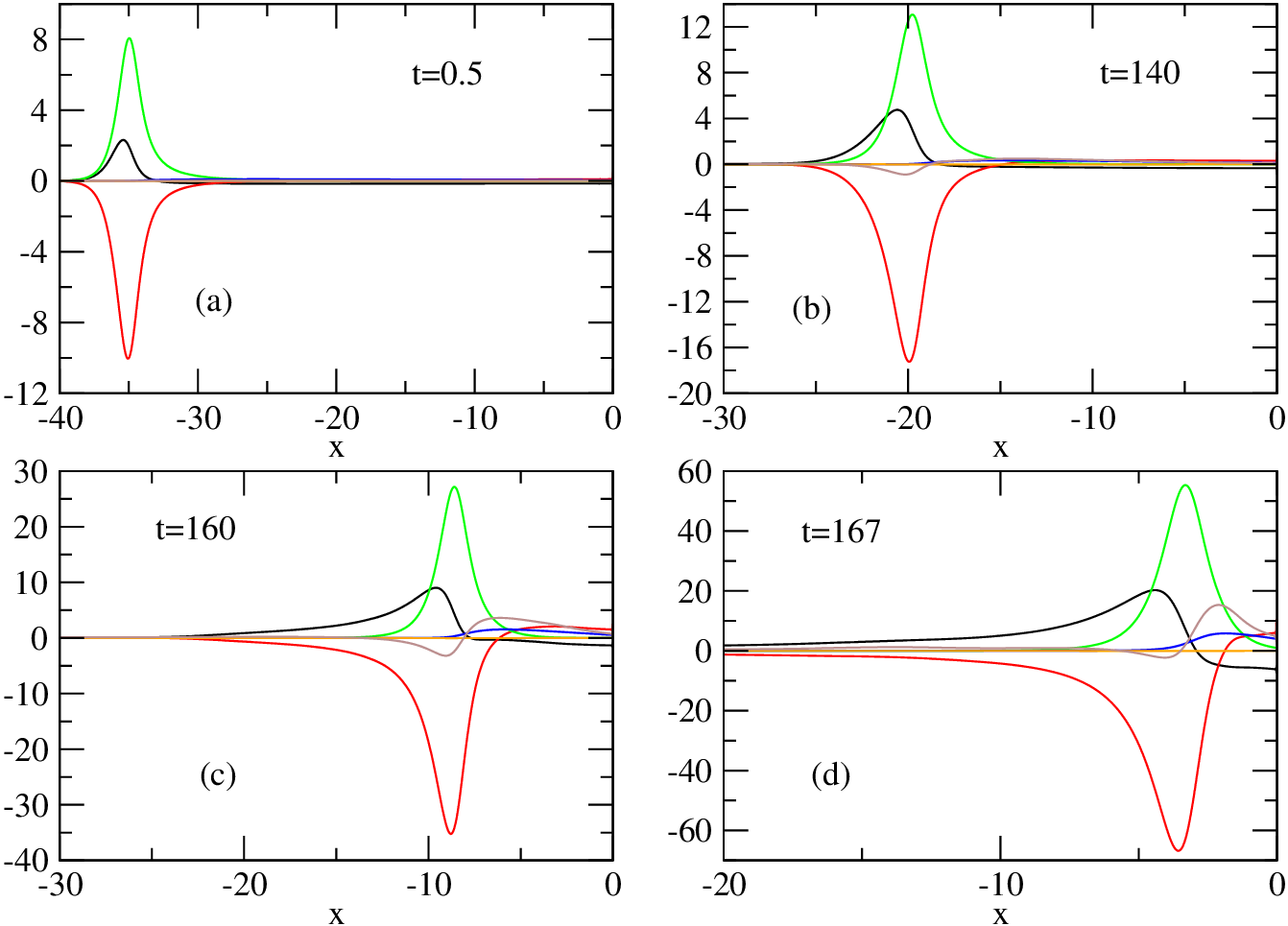}
\caption{Contributions of different terms in the pressure equation \eqref{eq:pressure} in the successful transition scenario at $D/D_{\textrm{CJ}}=0.3$: (1) - black; (2) - red; (3) - green; (4) - blue; (5) - orange; sum of all the terms - brown. The parameters are: $E=25$, $Q=50$, $\gamma=1.2$ and $c_f=0.012$. The computational domain size and resolution are $L=1000$ and $N_{1/2}=100$, respectively.
 }
\label{fig:transition-success}
\end{figure}

\begin{figure}
\centering
\psfrag{(a)}[l][Bl][0.8][0]{(a)}
\psfrag{(b)}[l][Bl][0.8][0]{(b)}
\psfrag{(c)}[l][Bl][0.8][0]{(c)}
\psfrag{(d)}[l][Bl][0.8][0]{(d)}
\psfrag{t}[l][Bl][0.8][0]{t}
\psfrag{x}[l][Bl][0.8][0]{$x$}
\psfrag{p}[l][Bl][0.8][0]{$p$}
\psfrag{u}[l][Bl][0.8][0]{$u$}
\psfrag{t=0.5}[l][Bl][0.8][0]{\hspace*{-0cm}$t=0.5$}
\psfrag{t=140}[l][Bl][0.8][0]{\hspace*{-1cm}$t=140$}
\psfrag{t=160}[l][Bl][0.8][0]{\hspace*{-1cm}$t=160$}
\psfrag{t=167}[l][Bl][0.8][0]{\hspace*{-0.5cm}$t=167$}
\psfrag{x}[l][Bl][0.8][0]{$x$}
\psfrag{10}[][][0.9][0]{$10$}
\psfrag{12}[][][0.9][0]{$12$}
\psfrag{15}[][][0.9][0]{$15$}
\psfrag{20}[][][0.9][0]{$20$}
\psfrag{30}[][][0.9][0]{$30$}
\psfrag{40}[][][0.9][0]{$40$}
\psfrag{50}[][][0.9][0]{$50$}
\psfrag{60}[][][0.9][0]{$60$}
\psfrag{-4}[][][0.9][0]{$-4$}
\psfrag{-8}[][][0.9][0]{$-8$}
\psfrag{-12}[][][0.9][0]{$-12$}
\psfrag{-16}[][][0.9][0]{$-16$}
\psfrag{0}[][][0.9][0]{$0$}
\psfrag{1}[][][0.9][0]{$1$}
\psfrag{2}[][][0.9][0]{$2$}
\psfrag{3}[][][0.9][0]{$3$}
\psfrag{4}[][][0.9][0]{$4$}
\psfrag{5}[][][0.9][0]{$5$}
\psfrag{-50}[][][0.9][0]{$-50$}
\psfrag{-10}[][][0.9][0]{$-10$}
\psfrag{-20}[][][0.9][0]{$-20$}
\psfrag{-30}[][][0.9][0]{$-30$}
\psfrag{-40}[][][0.9][0]{$-40$}
\includegraphics[scale=0.5,angle=0]{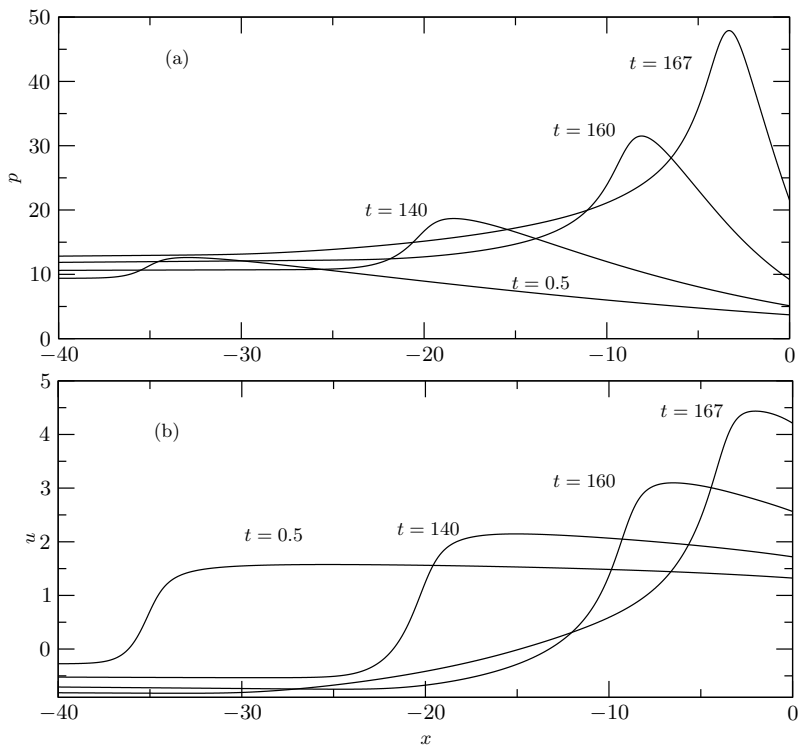}
\caption{Pressure and velocity profiles at different times corresponding to Fig. \ref{fig:transition-success}.
 }
\label{fig:transition-success-pres-temp}
\end{figure}

We now focus on a successful transition that starts at a set-valued solution at $D/D_{\textrm{CJ}}=0.3$, a case  chosen for its relatively low computational cost. We investigate the behaviour of all the terms in \eqref{eq:pressure} at four different times from the beginning of the simulation as displayed in Fig. \ref{fig:transition-success}. What is notable in this figure is that the dominant effects in the fire (by which we refer to the chemical heat release profile) are those of advection (1), dilatation (2) and heat release (3).  
At $t=0.5$, an essentially steady-state structure is seen wherein the positive contributions due to the heat release and advection are balanced by the negative contribution due to the gas expansion.  From approximately the middle of the fire  downstream, the advection term is positive because $D-u>0$ and $p_x>0$ (see Fig. \ref{fig:transition-success-pres-temp}), i.e., the high-pressure gas from the front-side of the fire is advected to the low-pressure region at the back. The dilatation term is negative there because of $u_{x}>0$. 

As time evolves, we observe that even though the same three effects remain dominant, some prominent changes to the profiles can be seen. Most interestingly, the chemical heat release profile advances upstream relative to that of the dilatation, that is, the expansion wave begins to lag behind the fire. As a result of this imbalance, a pressure build-up takes place in the fire region. The advective term (1) increases then on the back side of the fire, contributing more to balancing the expansion term than the heat release -- see Figs. \ref{fig:transition-success}(c,d). The gas expands locally leading to the pressure drop, but that pressure drop is compensated by its rise due to the high pressure gas advecting  from the fire region. Ahead of the fire, the dilatation term is seen to begin to rise to large positive values, indicating that the gas is getting compressed there. This is a consequence of the pressure waves propagating toward the shock from the fire region. The friction term is seen to increase between the shock and the front-side of the fire. Its contribution is now comparable to that of the compression. The advection term becomes negative in this region due to the change of the sign of the pressure gradient. Overall, on the front side of the fire, the heat release, friction and compressibility all contribute positively to the pressure rise, and dominate over the negative contribution of the advection term. The fastest growth of pressure is seen to take place at the front side of the fire where the friction effect is most significant. On the back-side of the fire, the pressure time-derivative becomes negative due to the strong expansion, which the weakening heat  release there and the pressure advection are unable to compensate.

To summarise, the successful transition is characterised by the dominant roles played by the heat release, pressure advection, compressibility and friction, and by the compression waves generated by the fire region that create favourable upstream conditions for the fire to move toward the shock relative to the expansion wave. 

\begin{figure}
\centering
\psfrag{(a)}[l][Bl][0.8][0]{(a)}
\psfrag{(b)}[l][Bl][0.8][0]{(b)}
\psfrag{(c)}[l][Bl][0.8][0]{(c)}
\psfrag{(d)}[l][Bl][0.8][0]{(d)}
\psfrag{t=200}[l][Bl][0.8][0]{$t=200$}
\psfrag{t=2000}[l][Bl][0.8][0]{$t=2000$}
\psfrag{t=4000}[l][Bl][0.8][0]{$t=4000$}
\psfrag{t=8000}[l][Bl][0.8][0]{$t=8000$}
\psfrag{x}[l][Bl][0.8][0]{$x$}
\psfrag{0}[][][0.9][0]{$0$}
\psfrag{-4}[][][0.9][0]{$-4$}
\psfrag{-8}[][][0.9][0]{$-8$}
\psfrag{4}[][][0.9][0]{$4$}
\psfrag{8}[][][0.9][0]{$8$}
\psfrag{1}[][][0.9][0]{$1$}
\psfrag{-50}[][][0.9][0]{$-50$}
\psfrag{-10}[][][0.9][0]{$-10$}
\psfrag{-20}[][][0.9][0]{$-20$}
\psfrag{-30}[][][0.9][0]{$-30$}
\psfrag{-40}[][][0.9][0]{$-40$}
\psfrag{-50}[][][0.9][0]{$-50$}
\psfrag{-60}[][][0.9][0]{$-60$}
\psfrag{-80}[][][0.9][0]{$-80$}
\psfrag{-100}[][][0.9][0]{$-100$}
\psfrag{-680}[][][0.9][0]{$-680$}
\psfrag{-690}[][][0.9][0]{$-690$}
\psfrag{-700}[][][0.9][0]{$-700$}
\includegraphics[scale=0.5,angle=0]{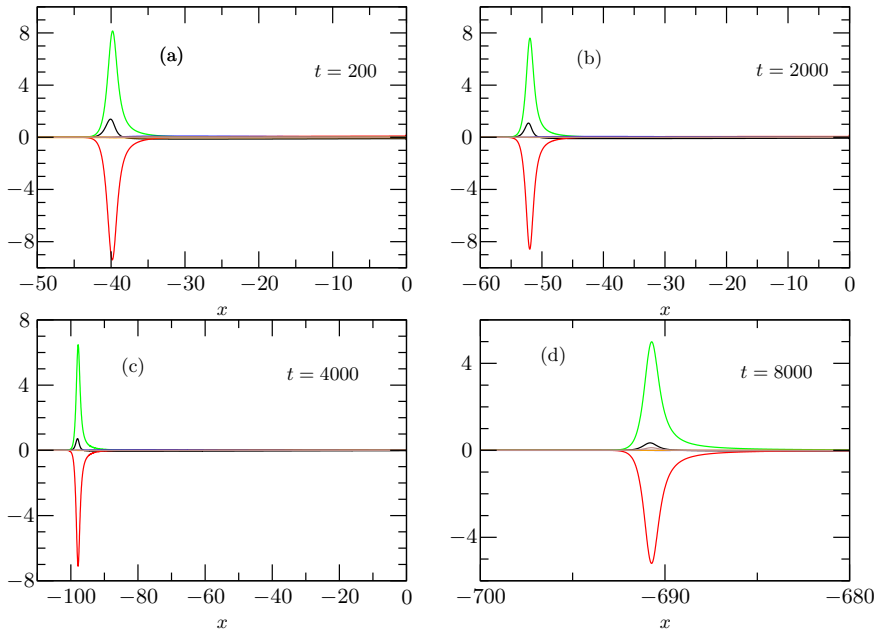}
\caption{Contributions of different terms in the pressure equation \eqref{eq:pressure} in the failing scenario (point $g$): (1) - black; (2) - red; (3) - green; (4) - blue; (5) - orange; sum of all the terms - brown. The parameters are: $E=25$, $Q=50$, $\gamma=1.2$ and $c_f=0.012$. The computational domain size and resolution are $L=1000$ and $N_{1/2}=100$, respectively.
 }
\label{fig:transition-failure}
\end{figure}

In the failing scenario shown in Fig. \ref{fig:transition-failure}, it is still heat release, expansion and advection that are the dominant players, however the fire is unable to move upstream relative to the position of the expansion wave. The fire is swept downstream with diminishing magnitudes of all the effects that contribute to the pressure rise. There is no noticeable pressure increase ahead of the fire.

Lastly, we point out  that the heat-loss term is seen to be negligible in the fire region compared to all the other terms in \eqref{eq:pressure}. The heat loss is however dominant downstream in the velocity-relaxation zone and is crucial due to its importance in establishing the steady-state equilibrium for the low velocity solutions. Without the heat loss, there would be no set-valued steady state solutions, hence no slow transient evolution at the early times in the transition seen in Fig. \ref{fig:E25-c-h}(b).

\section{\label{sec:Conclusions}Discussion and conclusions}

In this work, we analyze the stability properties of the set-valued solutions for gaseous detonations in systems with losses of momentum and heat. The peculiar nature of the steady-state solutions identified in \cite{semenko2016} indicates that low velocity one-dimensional detonations may be able to propagate in a nearly steady manner over long distances before transitioning to high-velocity regimes, predictions that we confirmed by the calculations made here. Thus, the existence of set-valued solutions points to a possible stabilization mechanism for low velocity detonations observed in experiments. We find that solutions at low velocities are neutrally stable because they are not isolated. Any particular steady-state solution is surrounded by a continuum of other steady-state solutions. Therefore, perturbations of such solutions do not lead to strong instability, but rather lead to a slow continuous passage of the solution through the steady states in the set-valued region. For this reason, transition to the high-velocity regimes or failure of the wave are observed to take place over long time periods, during which the waves evolve in a nearly steady non-oscillatory fashion. 

The early transients for the set-valued solutions which appear  nearly steady have evolution time scales that are comparable to the particle passage times across the reactions zone of the corresponding steady-state solutions. Those time scale are much larger than the time scales of the ideal detonations at the same mixture parameters. The latter are the scales that are used in our computations. Thus, even though the early transients appear quasi-steady on the chosen  $O(1)$ time scale of the ideal detonation, they are not quasi-steady on the time scale of the corresponding non-ideal solution. For this reason, we do not use the term ``quasi-steady'' in referring to these solutions. Nevertheless, the ideal time scales are the appropriate reference scale for all the solutions, and the low velocity waves do indeed appear nearly steady on those scales. One could draw a parallel with an actual physical experiment in which the velocities of the LVD are compared with the ideal velocities with both measured on the same time scale.

The existence of the large set of solutions at low velocities implies a certain degree of indeterminacy -- there appears to be no simple mechanism that selects a particular solution in the set-valued region. In other words, nearly steady regimes in a wide range of velocities are possible depending on particular initial conditions. As explained in \cite{semenko2016}, the existence of the set-valued solutions is tied to the indeterminacy of the boundary conditions at infinity. Once the temperature or pressure at infinity is specified, a particular $D$-$c_{f}$ curve is selected from the infinite set. Therefore, the choice of a particular solution for the low velocity detonation depends on the conditions far downstream.

The previous discussion is consistent with the experimental observations, which have shown that waves propagating in a continuous range of velocities
between the ambient sound speed and the ideal velocity are often observed \citep{lyamin1991propagation}. Our simulations indicate similar behavior:
nearly steady solutions exist at all velocities within the set-valued range that correspond to the sonic-velocity regimes or the low velocity detonation regimes. At the lower end of the velocity range, frictional heating of the gas is responsible for sustaining the reaction. As the velocity increases, so does the role of shock heating relative to frictional heating. On the upper branch of the $D$-$c_{f}$ curve, above the set-valued region (i.e., in the ``quasi-detonation'' regime),
the shock heating is dominant. These conclusions are consistent with several existing proposals on the mechanisms of such detonations \citep{korzhavin1999one,babkin1991propagation,babkin2012fast}.

Our present findings suggest a stabilization mechanism to the unconditionally unstable behavior of solutions near the middle branch of a typical
reverse $S$ shaped response curve that is found in the absence of heat losses. The presence of heat losses is then a key to this mechanism,
as no set-valued solutions exist without heat losses. We should emphasize that the low velocity regimes we computed here are not steady-state
stable solutions. They are nearly steady long-lived neutrally stable regimes that eventually fail or transition to high velocity regimes.

All the preceding conclusions apply to smaller activation energies as well. We observe qualitatively the same dynamics. At large
activation energies, we find that detonation often fails, even when starting the solution on the top branch. 
We remark finally that in view of the simplified nature of the modeling employed here as far as the reaction rate and loss factors, our predictions
are primarily qualitative in nature. It is therefore important to investigate how these predictions can be extended and compared with
available experiments at a quantitative level. Our preliminary calculations indicate that for the quantitative agreement,  modifications
are necessary as to the specific forms of the heat release function, loss terms and mixture thermodynamics \citep{Kasimov-Semenko-2016}.
However, the one-dimensional modeling with more complex physics still yields set-valued solutions, and the computed $D$-$c_{f}$ dependencies
are found to be closer to experimental data with an appropriate choice of the modeling parameters. We intend to continue our exploration
of the underlying mechanisms of low velocity solutions with such complex modeling and hope to report our findings in a follow-up work.

\section*{Acknowledgments}

Research reported in this publication was supported by the King Abdullah
University of Science and Technology (KAUST).

\bibliography{/Users/aslankasimov/Dropbox/Biblioteka/akasimov-refs}

\begin{thebibliography}{36}
\expandafter\ifx\csname natexlab\endcsname\relax\def\natexlab#1{#1}\fi
\def\au#1{#1} \def\ed#1{#1} \def\yr#1{#1}\def\at#1{#1}\def\jt#1{\textit{#1}}
  \def\bt#1{#1}\def\bvol#1{\textbf{#1}} \def\vol#1{#1} \def\pg#1{#1}
  \def\publ#1{#1}\def\arxiv#1{#1}\def\org#1{#1}\def\st#1{\textit{#1}}

\bibitem[Babkin(2012)]{babkin2012fast}
{\sc \au{Babkin, V.~S.}} \yr{2012}  \at{Fast gas combustion in systems with
  hydraulic resistance}.  \jt{Combustion, Explosion, and Shock Waves}
  \bvol{48}~(3),  \pg{278--287}.

\bibitem[Babkin {\em et~al.\/}(1991)Babkin, Korzhavin \&
  Bunev]{babkin1991propagation}
{\sc \au{Babkin, V.~S.}, \au{Korzhavin, A.~A.} \& \au{Bunev, V.~A.}} \yr{1991}
  \at{Propagation of premixed gaseous explosion flames in porous media}.
  \jt{Combustion and Flame}  \bvol{87}~(2),  \pg{182--190}.

\bibitem[Brailovsky {\em et~al.\/}(2012)Brailovsky, Kagan \&
  Sivashinsky]{brailovsky2012combustion}
{\sc \au{Brailovsky, I.}, \au{Kagan, L.} \& \au{Sivashinsky, G.}} \yr{2012}
  \at{Combustion waves in hydraulically resisted systems}.  \jt{Philosophical
  Transactions of the Royal Society A: Mathematical, Physical and Engineering
  Sciences}  \bvol{370}~(1960),  \pg{625--646}.

\bibitem[Brailovsky \& Sivashinsky(2000)]{brailovsky2000hydraulic-a}
{\sc \au{Brailovsky, I.} \& \au{Sivashinsky, G.}} \yr{2000}  \at{Hydraulic
  resistance and multiplicity of detonation regimes}.  \jt{Combustion and
  flame}  \bvol{122}~(1),  \pg{130--138}.

\bibitem[Brailovsky \& Sivashinsky(2002)]{brailovsky2002effects}
{\sc \au{Brailovsky, I.} \& \au{Sivashinsky, G.}} \yr{2002}  \at{Effects of
  momentum and heat losses on the multiplicity of detonation regimes}.
  \jt{Combustion and flame}  \bvol{128}~(1),  \pg{191--196}.

\bibitem[Camargo {\em et~al.\/}(2010)Camargo, Ng, Chao \&
  Lee]{camargo2010propagation}
{\sc \au{Camargo, A.}, \au{Ng, H.~D.}, \au{Chao, J.} \& \au{Lee, J. H.~S.}}
  \yr{2010}  \at{Propagation of near-limit gaseous detonations in small
  diameter tubes}.  \jt{Shock Waves}  \bvol{20}~(6),  \pg{499--508}.

\bibitem[Chan \& Greig(1989)]{chan1989structures}
{\sc \au{Chan, C.~K.} \& \au{Greig, D.~R.}} \yr{1989} The structures of fast
  deflagrations and quasi-detonations.  \bt{In {\em Symposium (International)
  on Combustion\/}}, ,  \vol{vol.~22},  \pg{pp. 1733--1739}. Elsevier.

\bibitem[Ciccarelli \& Dorofeev(2008)]{ciccarelli2008flame}
{\sc \au{Ciccarelli, G.} \& \au{Dorofeev, S.}} \yr{2008}  \at{Flame
  acceleration and transition to detonation in ducts}.  \jt{Progress in Energy
  and Combustion Science}  \bvol{34}~(4),  \pg{499--550}.

\bibitem[Dionne {\em et~al.\/}(2000)Dionne, Ng \& Lee]{dionne2000transient}
{\sc \au{Dionne, J.~P.}, \au{Ng, H.~D.} \& \au{Lee, J. H.~S.}} \yr{2000}
  \at{Transient development of friction-induced low-velocity detonations}.
  \jt{Proceedings of the Combustion Institute}  \bvol{28}~(1),  \pg{645--651}.

\bibitem[Gao {\em et~al.\/}(2016)Gao, Ng \& Lee]{gao2016near}
{\sc \au{Gao, Y.}, \au{Ng, H.~D.} \& \au{Lee, J. H.~S.}} \yr{2016}
  \at{Near-limit propagation of gaseous detonations in narrow annular
  channels}.  \jt{Shock Waves}  \pg{pp. 1--9}.

\bibitem[Gelfand {\em et~al.\/}(1991)Gelfand, Frolov \&
  Nettleton]{gelfand1991gaseous}
{\sc \au{Gelfand, B.~E.}, \au{Frolov, S.~M.} \& \au{Nettleton, M.~A.}}
  \yr{1991}  \at{Gaseous detonations--a selective review}.  \jt{Progress in
  energy and combustion science}  \bvol{17}~(4),  \pg{327--371}.

\bibitem[Henrick {\em et~al.\/}(2006)Henrick, Aslam \&
  Powers]{HenrickAslamPowers2006}
{\sc \au{Henrick, A.~K.}, \au{Aslam, T.~D.} \& \au{Powers, J.~M.}} \yr{2006}
  \at{Simulations of pulsating one-dimensional detonations with true fifth
  order accuracy}.  \jt{J. Comput. Phys.}  \bvol{213}~(1),  \pg{311--329}.

\bibitem[Higgins(2012)]{higgins2012steady}
{\sc \au{Higgins, A.~J.}} \yr{2012}  \at{Steady one-dimensional detonations}.
  \bt{In {\em Shock Waves Science and Technology Library, Vol. 6\/}},  \pg{pp.
  33--105}.  \publ{Springer}.

\bibitem[Ishii \& Monwar(2011)]{ishii2011detonation}
{\sc \au{Ishii, K.} \& \au{Monwar, M.}} \yr{2011}  \at{Detonation propagation
  with velocity deficits in narrow channels}.  \jt{Proceedings of the
  Combustion Institute}  \bvol{33}~(2),  \pg{2359--2366}.

\bibitem[Kasimov \& Semenko(2016)]{Kasimov-Semenko-2016}
{\sc \au{Kasimov, A.~R.} \& \au{Semenko, R.}} \yr{2016}  \at{On modeling
  gaseous detonation in a porous medium by one-dimensional {Euler} equations}.
  \jt{Combustion and Explosion}  \bvol{9}~(4),  \pg{19--26}.

\bibitem[Korzhavin {\em et~al.\/}(1999)Korzhavin, Bunev, Babkin, Lawes \&
  Bradley]{korzhavin1999one}
{\sc \au{Korzhavin, A.~A.}, \au{Bunev, V.~A.}, \au{Babkin, V.~S.}, \au{Lawes,
  M.} \& \au{Bradley, D.}} \yr{1999}  \at{On one regime of low-velocity
  detonation in porous media}.  \jt{Gaseous and Heterogeneous Detonations:
  Science to Applications, ENAS Publ., Moscow}  \pg{pp. 255--268}.

\bibitem[Lee \& Stewart(1990)]{LeeStewart90}
{\sc \au{Lee, H.~I.} \& \au{Stewart, D.~S.}} \yr{1990}  \at{Calculation of
  linear detonation instability: One-dimensional instability of plane
  detonation}.  \jt{J. Fluid Mech.}  \bvol{212},  \pg{103--132}.

\bibitem[Lee(2008)]{Lee-2008}
{\sc \au{Lee, J. H.~S.}} \yr{2008} {\em The Detonation Phenomenon\/}.
  \publ{Cambridge University Press}.

\bibitem[Lee {\em et~al.\/}(1985)Lee, Knystautas \& Chan]{lee1985turbulent}
{\sc \au{Lee, J. H.~S.}, \au{Knystautas, R.} \& \au{Chan, C.~K.}} \yr{1985}
  Turbulent flame propagation in obstacle-filled tubes.  \bt{In {\em Symposium
  (International) on Combustion\/}}, ,  \vol{vol.~20},  \pg{pp. 1663--1672}.
  Elsevier.

\bibitem[Lee {\em et~al.\/}(1984)Lee, Knystautas \& Freiman]{lee1984high}
{\sc \au{Lee, J. H.~S.}, \au{Knystautas, R.} \& \au{Freiman, A.}} \yr{1984}
  \at{{High speed turbulent deflagrations and transition to detonation in $H_2$
  - air mixtures}}.  \jt{Combustion and flame}  \bvol{56}~(2),  \pg{227--239}.

\bibitem[Lyamin {\em et~al.\/}(1991)Lyamin, Mitrofanov, Pinaev \&
  Subbotin]{lyamin1991propagation}
{\sc \au{Lyamin, G.~A.}, \au{Mitrofanov, V.~V.}, \au{Pinaev, A.~V.} \&
  \au{Subbotin, V.~A.}} \yr{1991}  \at{Propagation of gas explosion in channels
  with uneven walls and in porous media}.  \jt{Dynamic Structure of Detonation
  in Gaseous and Dispersed Media, Kluwer Academ., Netherlands}  \pg{pp.
  51--75}.

\bibitem[Lyamin \& Pinaev(1985)]{lyamin1985supersonic}
{\sc \au{Lyamin, G.~A.} \& \au{Pinaev, A.~V.}} \yr{1985} Supersonic
  (detonation) combustion of gases in inert porous media.  \bt{In {\em Soviet
  Physics Doklady\/}}, ,  \vol{vol.~30},  \pg{p. 694}.

\bibitem[Makris {\em et~al.\/}(1995)Makris, Shafique, Lee \&
  Knystautas]{makris1995influence}
{\sc \au{Makris, A.}, \au{Shafique, H.}, \au{Lee, J. H.~S.} \& \au{Knystautas,
  R.}} \yr{1995}  \at{Influence of mixture sensitivity and pore size on
  detonation velocities in porous media}.  \jt{Shock Waves}  \bvol{5}~(1-2),
  \pg{89--95}.

\bibitem[Manzhalei(1992)]{manzhalei1992detonation}
{\sc \au{Manzhalei, V.~I.}} \yr{1992}  \at{Detonation regimes of gases in
  capillaries}.  \jt{Combustion, Explosion, and Shock Waves}  \bvol{28}~(3),
  \pg{296--302}.

\bibitem[Manzhalei(1998)]{manzhalei1998gas}
{\sc \au{Manzhalei, V.~I.}} \yr{1998}  \at{Gas detonation in a flat channel of
  50-$\mu$m depth}.  \jt{Combustion, Explosion and Shock Waves}  \bvol{34}~(6),
   \pg{662--664}.

\bibitem[Manzhalei(1999)]{manzhalei1999low}
{\sc \au{Manzhalei, V.~I.}} \yr{1999}  \at{Low-velocity detonation limits of
  gaseous mixtures}.  \jt{Combustion, explosion, and shock waves}
  \bvol{35}~(3),  \pg{296--302}.

\bibitem[Radulescu \& Lee(2002)]{radulescu2002failure}
{\sc \au{Radulescu, M.I.} \& \au{Lee, J.H.S.}} \yr{2002}  \at{The failure
  mechanism of gaseous detonations: experiments in porous wall tubes}.
  \jt{Combustion and flame}  \bvol{131}~(1-2),  \pg{29--46}.

\bibitem[Schelkin(1949)]{schelkin1949fast}
{\sc \au{Schelkin, K.~I.}} \yr{1949} {\em Fast Combustion and Spin Detonation
  of Gases\/}.  \publ{Moscow: Voenizdat}.

\bibitem[Semenko {\em et~al.\/}(2016)Semenko, Faria, Kasimov \&
  Ermolaev]{semenko2016}
{\sc \au{Semenko, R.}, \au{Faria, L.~M.}, \au{Kasimov, A.~R.} \& \au{Ermolaev,
  B.~S.}} \yr{2016}  \at{Set-valued solutions for non-ideal detonation}.
  \jt{Shock Waves}  \bvol{26}~(2),  \pg{141--160}.

\bibitem[Sow {\em et~al.\/}(2014)Sow, Chinnayya \& Hadjadj]{sow2014mean}
{\sc \au{Sow, A.}, \au{Chinnayya, A.} \& \au{Hadjadj, A.}} \yr{2014}  \at{Mean
  structure of one-dimensional unstable detonations with friction}.
  \jt{Journal of Fluid Mechanics}  \bvol{743},  \pg{503--533}.

\bibitem[Teodorczyk {\em et~al.\/}(1989)Teodorczyk, Lee \&
  Knystautas]{teodorczyk1989propagation}
{\sc \au{Teodorczyk, A.}, \au{Lee, J. H.~S.} \& \au{Knystautas, R.}} \yr{1989}
  Propagation mechanism of quasi-detonations.  \bt{In {\em Symposium
  (International) on Combustion\/}}, ,  \vol{vol.~22},  \pg{pp. 1723--1731}.
  Elsevier.

\bibitem[Teodorczyk {\em et~al.\/}(1991)Teodorczyk, Lee \&
  Knystautas]{teodorczyk1991structure}
{\sc \au{Teodorczyk, A.}, \au{Lee, J. H.~S.} \& \au{Knystautas, R.}} \yr{1991}
  The structure of fast turbulent flames in very rough, obstacle-filled
  channels.  \bt{In {\em Symposium (International) on Combustion\/}}, ,
  \vol{vol.~23},  \pg{pp. 735--741}. Elsevier.

\bibitem[Zel'dovich(1940)]{Zeldovich1940}
{\sc \au{Zel'dovich, Y.~B.}} \yr{1940}  \at{On the theory of propagation of
  detonation in gaseous systems}.  \jt{J. Exp. Theor. Phys.}  \bvol{10}~(5),
  \pg{542--569}.

\bibitem[Zel'dovich {\em et~al.\/}(1987)Zel'dovich, Gel'fand, Kazhdan \&
  Frolov]{zel1987detonation}
{\sc \au{Zel'dovich, Y.~B.}, \au{Gel'fand, B.~E.}, \au{Kazhdan, Y.~M.} \&
  \au{Frolov, S.~M.}} \yr{1987}  \at{Detonation propagation in a rough tube
  taking account of deceleration and heat transfer}.  \jt{Combustion,
  Explosion, and Shock Waves}  \bvol{23}~(3),  \pg{342--349}.

\bibitem[Zel'dovich \& Kompaneets(1960)]{zeldovich1960theory}
{\sc \au{Zel'dovich, Y.~B.} \& \au{Kompaneets, A.~S.}} \yr{1960} {\em Theory of
  detonation\/}.  \publ{New York: Academic Press}.

\bibitem[Zhang \& Lee(1994)]{zhang1994friction}
{\sc \au{Zhang, F.} \& \au{Lee, J. H.~S.}} \yr{1994}  \at{Friction-induced
  oscillatory behaviour of one-dimensional detonations}.  \jt{Proceedings of
  the Royal Society of London. Series A: Mathematical and Physical Sciences}
  \bvol{446}~(1926),  \pg{87--105}.

\end{thebibliography}
\bibliographystyle{jfm}

\end{document}